\documentclass[amsmath,amssymb,prb,floatfix,twocolumn]{revtex4}

\usepackage{graphicx}
\usepackage{dcolumn}
\usepackage{bm}
\usepackage{times}
\usepackage[usenames,dvipsnames]{color}
\usepackage{colordvi,epsf}

\begin{document}

\title{Enhanced skyrmion stability due to exchange frustration}
\author{S.~von Malottki${}^1$, B.~Dup\'e${}^{1,2}$, P.~F.~Bessarab${}^{3,4}$, A.~Delin${}^{5,6}$ and S.~Heinze${}^{1}$}
\affiliation{${}^1$Institute of Theoretical Physics and Astrophysics, University of Kiel,
24098 Kiel, Germany}
\affiliation{${}^2$Institute of Physics, University of Mainz, 
55128 Mainz, Germany}
\affiliation{${}^3$School of Engineering and Natural Sciences - Science Institute, University of Iceland,
107 Reykjavik, Iceland}
\affiliation{${}^4$University ITMO, St.~Petersburg 197101, Russia}
\affiliation{${}^5$Department of Applied Physics, School of Engineering Sciences,
KTH Royal Institute of Technology, Electrum 229, SE-16440 Kista, Sweden}
\affiliation{${}^6 $Department of Physics and Astronomy, Materials Theory Division, Uppsala University,
Box 516, SE-75120 Uppsala, Sweden}

\date{\today}

\begin{abstract}
\noindent
Skyrmions are localized, topologically non-trivial spin structures which have raised high hopes for future spintronic applications.  
A key issue is skyrmion stability with respect to annihilation into the ferromagnetic state. 
Energy barriers for this collapse have been calculated taking only nearest neighbor exchange interactions into account.
Here, we demonstrate that exchange interactions beyond nearest neighbors 
can be essential to describe stability of skyrmionic spin structures. 
We focus on the prototypical film system Pd/Fe/Ir(111) and 
demonstrate that an effective nearest-neighbor exchange or micromagnetic model can only 
account for equilibrium properties such as the skyrmion profile or the 
zero temperature phase diagram. However, energy barriers and critical fields
of skyrmion collapse as well as skyrmion lifetimes are drastically underestimated 
since the energy of the transition state 
cannot be accurately described. Antiskyrmions are not even metastable. 
Our work shows that frustration of exchange interactions is a route
towards enhanced skyrmion stability even in systems with a ferromagnetic ground state.
\end{abstract}

      
\maketitle

Skyrmions have been predicted to occur in magnetic materials based on micromagnetic models more than 25 years ago \cite{Bogdanov1989,Bogdanov1994}. 
Experimentally skyrmions were first observed in bulk materials with a chiral crystal structure \cite{Muehlbauer:09.1,Yu:10.1,Seki2012}. In these systems 
skyrmions are stabilized by the Dzyaloshinskii-Moriya interaction (DMI) \cite{Moriya,Dzyaloshinskii}.
The DMI occurs due to spin-orbit coupling (SOC) in systems with a broken structural inversion symmetry. Therefore, it is present also
at surfaces or interfaces \cite{Crepieux1998,Bode2007} and can induce spin spirals \cite{Bode2007,Ferriani2008}, chiral domain walls 
\cite{Heide2008,Meckler2009,nmat3675,Ryu2013,thiaville2012dynamics} 
and skyrmions \cite{Heinze:11.1,Romming:13.1}. Since transition-metal interfaces and multilayers are at the heart of technologically
established spintronic devices such as read heads of hard-disk drives based on the giant magnetoresistance effect \cite{PhysRevLett.61.2472,PhysRevB.39.4828}
the discovery of skyrmions at interfaces \cite{Heinze:11.1,Romming:13.1} sparked great interest in using them for novel device concepts \cite{Fert:13.1,Sampaio:13.1}.

For potential applications \cite{Kiselev2011,Fert:13.1,Sampaio:13.1,Nagaosa:13.1} the stability of magnetic skyrmions becomes a crucial issue. 
An experimentally well studied system for nanoscale skyrmions at interfaces is Pd/Fe/Ir(111) \cite{Romming:13.1,Romming:15.1,Hanneken:15.1,Hagemeister:15.1,Leonov:16.1}, 
i.e.~a single atomic layer of Pd grown on an Fe monolayer 
on the Ir(111) surface. Based on fitting field dependent experimental skyrmion profiles to the micromagnetic model effective parameters have been 
obtained \cite{Romming:15.1} which were used to discuss properties of isolated skyrmions \cite{Leonov:16.1}.
Current-induced skyrmion annihilation has also been studied combining the experiments with
Monte-Carlo simulations from which energy barriers of about 50~meV have been estimated \cite{Hagemeister:15.1}. 
Energy barriers protecting skyrmion states in ultrathin films have been obtained by calculating minimum energy paths (MEPs) for the skyrmion collapse into the ferromagnetic state 
\cite{Bessarab:15.1,Rohart:16.1,Lobanov:16.1,Uzdin:17.1}
which due to the huge configuration space is a non-trivial task \cite{Bessarab:17.1,Rohart:17.1}.

While an atomistic spin model has been used in these
approaches, the exchange interaction has been treated within the nearest-neighbor (NN) approximation which corresponds to the exchange stiffness 
within the micromagnetic model.  
For Pd/Fe/Ir(111) it has been shown based on density functional theory (DFT) calculations
that the exchange interaction is ferromagnetic for NN while it is antiferromagnetic for the 2nd and 3rd NN \cite{Dupe:14.1,Simon:14.1}. 
Therefore, it is not clear in how far spin models using an effective NN exchange interaction are sufficient to describe skyrmion properties in these systems.

Here, we demonstrate that frustrated exchange greatly enhances skyrmion stability in ultrathin films and can lead to metastable antiskyrmions. We use
atomistic spin dynamics simulations parametrized from DFT for Pd/Fe/Ir(111) and obtain energy barriers for skyrmion annihilation into the ferromagnetic state by 
applying the geodesic nudged elastic band (GNEB) approach \cite{Bessarab:15.1}. We compare these calculations with an effective NN exchange model
based on our DFT calculations which yields parameters  
that are in good agreement with those obtained from experiments \cite{Romming:15.1,Hagemeister:15.1}. We find that
equilibrium skyrmion properties such as field-dependent profiles or the zero temperature phase diagram are well described by
the effective NN exchange approach. However, energy barriers of skyrmion collapse and skyrmion lifetimes
are greatly underestimated if the frustrated exchange interaction is
mapped to an effective NN exchange. Antiskyrmions are not even metastable in the effective NN exchange model. 

\begin{figure*}[thp]
\centering
\includegraphics[width=0.92\textwidth]{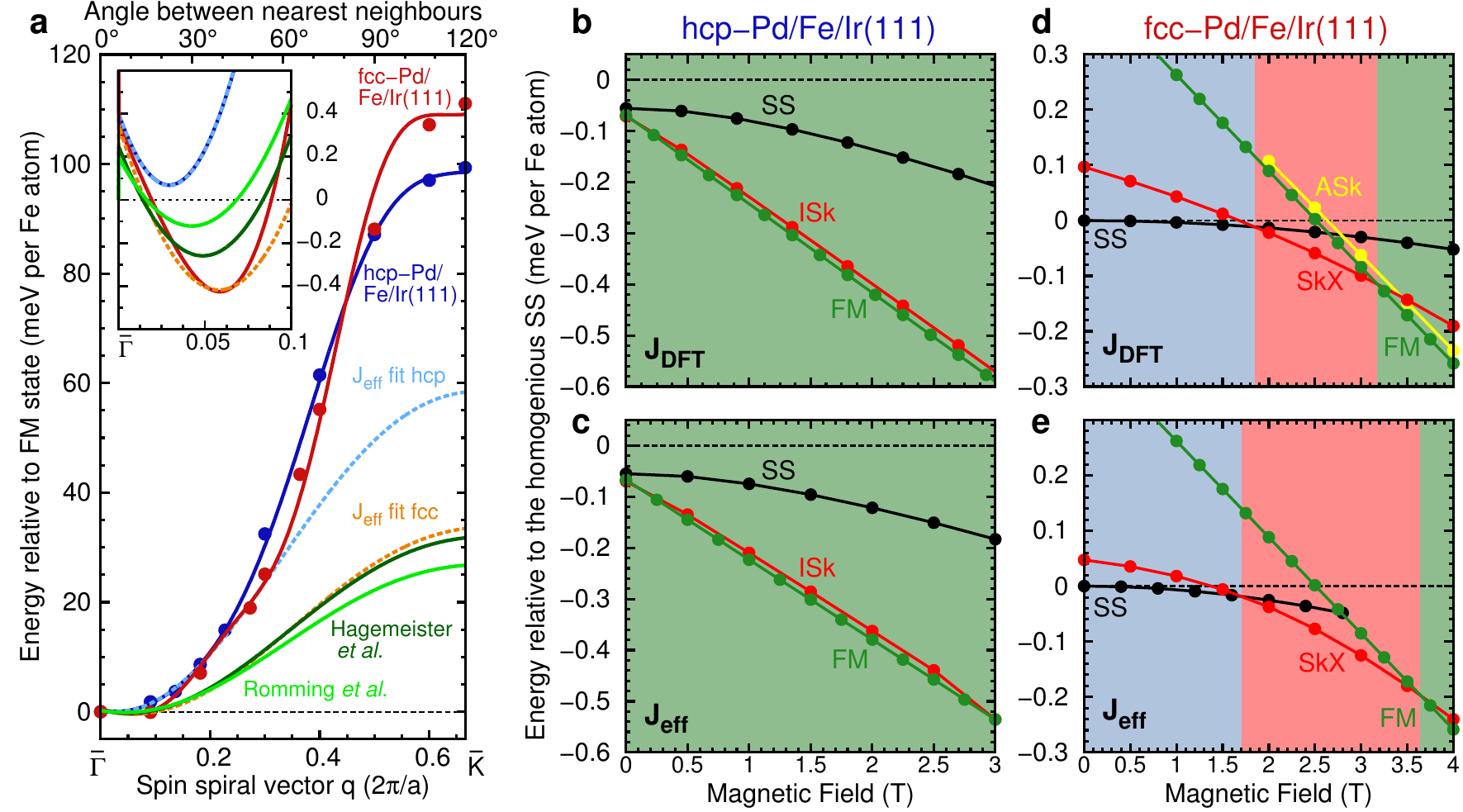}
\caption{
\textbf{Energy dispersion of spin spirals and zero temperature phase diagrams of Pd/Fe/Ir(111).}
\textbf{(a)} 
Energy dispersion of homogeneous spin spirals for Pd/Fe/Ir(111) as a function of the spin spiral vector \textbf{q} along the $\bar{\Gamma}-\bar{\rm K}$-direction. 
The inset shows a zoom of the dispersion around the energy minima. Note that there is an offset for all curves at $\mathbf{q}=0$ due to the 
magnetocrystalline anisotropy which leads to an energy increase for all spin spiral states of $K/2$ with respect to the ferromagnetic state.
The filled circles 
are total energies obtained from DFT for fcc (red) or hcp (blue) stacking of the Pd layer including spin-orbit coupling. 
The solid lines are fits to the Heisenberg model including the Dzyaloshinskii-Moriya interaction.
The dashed lines are fits to the DFT energy dispersion 
close to the energy minima with an effective nearest-neighbor exchange interaction $J_{\rm eff}$ 
and effective DMI for fcc (orange) and hcp (blue) stacking.
For comparison, the energy dispersions for parameters given by Romming \textit{et al}. \cite{Romming:15.1} (light green) and Hagemeister \textit{et al.} \cite{Hagemeister:15.1}. (dark green) are shown. 
\textbf{(b,c)} Zero temperature phase diagram for hcp-Pd/Fe/Ir(111) obtained with the DFT ($J_{\rm DFT}$) and effective ($J_{\rm eff}$) parameters, respectively. 
The energies of the ferromagnetic (FM), isolated skyrmion (ISk) and relaxed spin spiral (SS) state are shown relative to the homogeneous spin spiral (dashed line).
The green color indicates the regime of the FM ground state.
\textbf{(d,e)} Zero temperature phase diagram for fcc-Pd/Fe/Ir(111)
as in \textbf{(b,c)} including the skyrmion lattice (SkX) and the isolated antiskyrmion (ASk) state.
Blue, red, and green color represents the regime of the SS, SkX, and FM ground state, respectively.
}
\label{fig:dispersion_stability}
\end{figure*}

\noindent {\textbf{Results}}

\noindent
\textbf{Atomistic spin model.} 
We describe the magnetic properties of Pd/Fe/Ir(111) using the spin Hamiltonian given by
\begin{eqnarray}
H = & - & \sum_{ij} J_{ij} (\mathbf m_i \cdot \mathbf m_j)
      -   \sum_{ij} \mathbf D_{ij} \cdot (\mathbf m_i \times \mathbf m_j) \nonumber  \\
    & - & \sum_i K (m_i^z)^2
      -   \sum_i \mu_s \mathbf B \cdot \mathbf m_i \, ,
\label{eq:modelH}
\end{eqnarray}
which describes the magnetic interactions between the magnetic moments $\mathbf{M}_i$ of
atoms at sites $\mathbf{R}_i$ where $\mathbf{m}_i=\mathbf{M}_i/ M_i$. 
The parameters for the exchange interaction ($J_{ij}$), the DMI ($\mathbf{D}_{ij}$), the magnetic moments ($\mu_s$) as well as an uniaxial
magnetocrystalline anisotropy ($K$) were obtained from DFT using the {\tt FLEUR} code \cite{Kurz-SS,Heide-DMI,Dupe:14.1,Zimmermann2014} (see methods for details).
Note that $J_{ij}$ and $\mathbf{D}_{ij}$ are defined per spin because each pairwise interaction contributes twice in the total energy. 
We do not explicitly include dipole-dipole interactions. However, for ultrathin films this energy term is very small -- on the order of 0.1~meV/atom --
and it can be effectively included into the magnetocrystalline anisotropy energy \cite{Draaisma:88.1,Lobanov:16.1}.

While the Fe layer follows the fcc stacking of the Ir(111) surface, both Pd stackings have been observed experimentally \cite{Kubetzka:17.1}.
Fig.~\ref{fig:dispersion_stability}a shows the energy dispersions $E({\mathbf q})$ of flat homogeneous spin spirals including the effect of SOC 
calculated from DFT for both hcp and fcc stacking of the Pd overlayer. 
A spin spiral is characterized by a wave vector ${\mathbf q}$ from the two-dimensional Brillouin zone (2D-BZ) and the magnetic moment of an atom at site ${\mathbf R}_i$ is given by 
$\mathbf{M}_i=M (\sin{({\mathbf q} {\mathbf R}_i)}⁡,\cos{({\mathbf q} {\mathbf R}_i)},0)$ with the size of the magnetic moment $M$. 
We find a value of about 2.7 $\mu_{\rm B}$ per Fe-atom in our calculations which is fairly constant for all ${\mathbf q}$ vectors. Combined with the magnetic moment of 0.3 $\mu_{\rm B}$ due to spin polarization of the Pd-atoms, we obtain a total moment of 3.0 $\mu_{\rm B}$ that we use for our simulations.
Fitting the energy dispersions along the high symmetry directions of the 2D-BZ using the first two terms of Eq.~(\ref{eq:modelH}) allows to extract the exchange constants and the DMI 
(see table \ref{tab:DFT_Jij}). Our computed shell resolved $J_{ij}$ agree well with those presented in Ref.~\onlinecite{Simon:14.1}.
In order to obtain a good fit of the DFT 
energy dispersion many shells need to be taken into account for the exchange while the DMI has been treated in the NN approximation. 

\begin{table*}[htb]
\centering
\begin{ruledtabular}
\begin{tabular*}{\hsize}{c@{\extracolsep{0ptplus1fil}}ccccccccccc}
{} & $J_1$ & $J_2$ & $J_3$ & $J_4$ & $J_5$ & $J_6$ & $J_7$ & $J_8$ & $J_9$ & $D_1$ & $K$ \\ 
hcp & 13.66 & $-0.51$ & $-2.88$ & 0.07 & 0.55 & - & - & - & - & 1.2 & 0.8\\ 
fcc & 14.40   & $-2.48$ & $-2.69$ & 0.52 & 0.74 & 0.28 & 0.16 & $-0.57$ & $-0.21$ & 1.0 & 0.7\\ 
\end{tabular*}
\end{ruledtabular}
\caption{Exchange constants, DMI and magnetocrystalline anisotropy obtained from DFT for hcp-Pd/Fe/Ir(111) and fcc-Pd/Fe/Ir(111). The positive sign of the DMI
         indicates favoring of right rotating spin spirals and the positive sign of $K$ indicates an easy out-of-plane magnetization direction.
         The coefficients are given in meV.
        }
\label{tab:DFT_Jij}
\end{table*}

\begin{table}[htb]
\centering
\begin{ruledtabular}
\begin{tabular*}{\hsize}{c@{\extracolsep{0ptplus1fil}}cccc}
{} & $J_{\rm eff}$ & $D_{\rm eff}$ & $K$ \\  
hcp fit & 6.44 & 1.2 & 0.8 \\
fcc fit & 3.68 & 1.39 & 0.7 \\ 
Ref.~\onlinecite{Romming:15.1} & 2.95 & 0.8 & 0.4 \\
Ref.~\onlinecite{Hagemeister:15.1} & 3.5 & 1.1 & 0.5 \\
\end{tabular*}
\end{ruledtabular}
\caption{Parameters of the effective NN exchange model obtained by fitting the DFT results for hcp-Pd/Fe/Ir(111) and fcc-Pd/Fe/Ir(111) 
         and values from the literature. For hcp $D_{\rm eff}$ has been chosen as the NN DFT value (cf.~table \ref{tab:DFT_Jij}).
         Sign convention as in table \ref{tab:DFT_Jij}. The coefficients are given in meV.
        }
\label{tab:J_eff}
\end{table}

It has been suggested to characterize the behavior of the exchange interaction close to the $\bar{\Gamma}$ point of the 2D-BZ -- corresponding to the ferromagnetic (FM) state -- by an effective 
NN exchange constant $J_{\rm eff}$ \cite{Dupe:16.1} --
an approach which has also been used in previous theoretical studies of skyrmion stability \cite{Bessarab:15.1,Hagemeister:15.1,Leonov:16.1}. 

From our DFT energy dispersion we can obtain $J_{\rm eff}$ 
by using only the NN terms of the first two terms in the spin Hamiltonian, Eq.~(\ref{eq:modelH}),
which yields along the $\bar{\Gamma}-\bar{\rm K}$ direction 
$E(q)=J_{\rm eff} \left( 2 \cos(q) + 4 \cos(q/2) \right)$ (with $q$ in units of $2 \pi/a$, see table \ref{tab:J_eff} for $J_{\rm eff}$).
We obtain an excellent approximation
around the local spin spiral energy minimum (cf.~inset of Fig.~\ref{fig:dispersion_stability}a). 
However, the fit drastically underestimates $E(q)$ from DFT for larger values of $q$. 
For fcc-Pd/Fe/Ir(111) we had to fit also the DMI to obtain a good agreement.
For spin spirals along $\bar{\Gamma}-\bar{\rm K}$ we use
$E(q)=J_{\rm eff} \left( 2 \cos(q) + 4 \cos(q/2) \right)+ D_{\rm eff} \left( 2 \sin(q) + 2 \sin (q/2) \right)$
and a fit interval close to the energy minimum (see table \ref{tab:J_eff} for $J_{\rm eff}$ and $D_{\rm eff}$).  

Effective NN exchange and DMI for Pd/Fe/Ir(111)
have previously been extracted by fitting experimental data to micromagnetic and Monte-Carlo simulations \cite{Romming:15.1,Hagemeister:15.1} 
(for values see table \ref{tab:J_eff}). The energy dispersions obtained with these values are displayed in Fig.~\ref{fig:dispersion_stability}a. 
These curves are between the DFT dispersions for hcp and fcc Pd stacking around the energy minimum (cf.~inset of Fig.~\ref{fig:dispersion_stability}a). 
For larger values of $q$, i.e.~larger angles between adjacent spins, they are close to the effective fit in fcc stacking but deviate 
drastically from the DFT curves. 

\begin{figure}[thp]
\centering
\includegraphics[trim= 0.0cm 0.0cm 1.2cm 0.0cm,clip,width=0.46\textwidth]{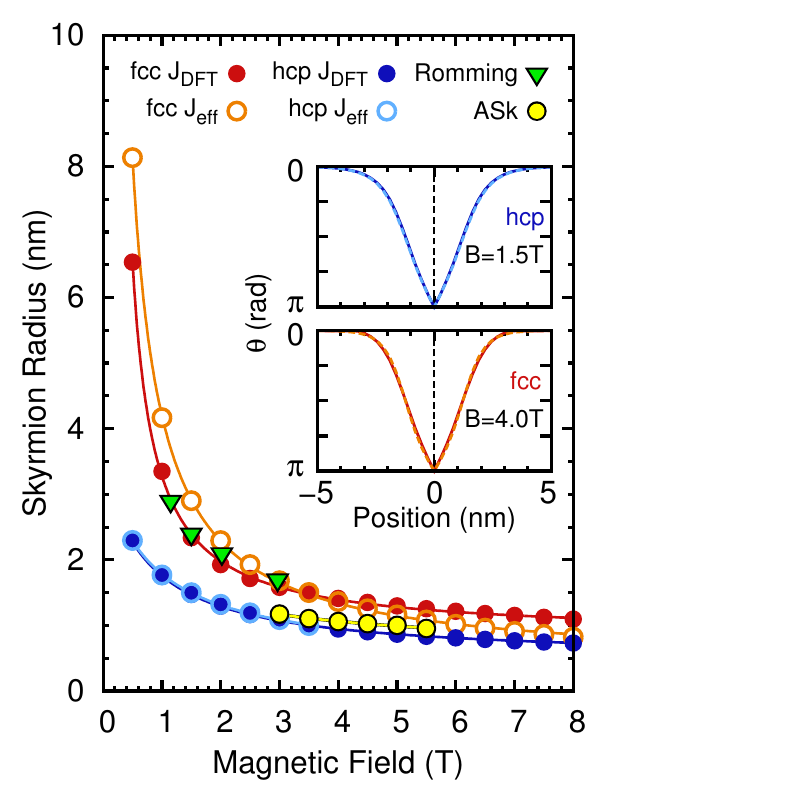}
\caption{
\textbf{Skyrmion radius vs.~magnetic field.} Radii of skyrmions in Pd/Fe/Ir(111) obtained for different parameter sets as a function of the 
applied magnetic field. As a reference, the radii obtained experimentally by Romming \textit{et al.} \cite{Romming:15.1} are shown as green triangles. 
Antiskyrmions (ASk) were only metastable for fcc-Pd/Fe/Ir(111) with DFT parameters.
}
\label{fig:radius}
\end{figure}

\noindent \textbf{Zero temperature phase diagrams.} To compare the effective description of the exchange (denoted as $J_{\rm eff}$) and DMI (for fcc-Pd/Fe/Ir(111)) 
and the full set of exchange constants from DFT 
(denoted as $J_{\rm DFT}$), we have obtained the phase diagram at zero temperature as a function of the applied magnetic field (Fig.~\ref{fig:dispersion_stability}b-e) 
using spin-dynamics simulations (see methods). Note that we have used the magnetocrystalline anisotropy obtained from DFT in both cases.

For hcp-Pd/Fe/Ir(111) the phase diagrams are very similar for both ways of treating the exchange [Fig.~\ref{fig:dispersion_stability}b,c]. 
The FM state is the ground state over the whole range of magnetic field values. Homogeneous spin spirals 
relax into a domain wall structure which gains energy due to DMI \cite{Heide2008}.  
Isolated skyrmions are metastable for both sets of parameters when an external field of at least $\approx 0.5 \mathrm{T}$ is applied. For smaller fields, the skyrmions become large bubbles 
with a fixed rotational sense of the domain walls surrounding the FM core. 

For fcc-Pd/Fe/Ir(111) [Fig.~\ref{fig:dispersion_stability}d,e]
the ground state is a spin spiral consistent with the energy minimum in the spin spiral dispersion curve (cf.~inset of Fig.~\ref{fig:dispersion_stability}a).
For both $J_{\rm eff}$ and $J_{\rm DFT}$ a skyrmion lattice is energetically favorable at a critical field and at even larger fields there is a transition to the FM 
phase. However, the critical fields at which these transitions occur are slightly different ($B_{\rm c}^{\rm eff} \approx 3.6$~T and $B_{\rm c}^{\rm DFT} \approx 3.2$~T) due to the
larger DMI obtained by the fit. Using the full set of DFT parameters we were also able to metastabilize antiskyrmions within the skyrmion lattice and FM phase.

\noindent
\textbf{Skyrmion radius.} In the FM phase isolated skyrmions are metastable in our spin dynamics simulations up to a critical field for both hcp-Pd/Fe/Ir(111) and fcc-Pd/Fe/Ir(111).
The skyrmion profiles were obtained by imposing the theoretical profile~\cite{Bogdanov-1994aa} and relaxing this spin structure within our spin dynamics simulation. 
The radii were extracted as in Ref.~\onlinecite{Bogdanov-1994aa}. The obtained relaxed skyrmion profiles for both descriptions of the exchange are nearly indistinguishable as 
exemplified in the insets of Fig.~\ref{fig:radius}.
The resulting radii decrease rapidly with applied magnetic field (cf.~Fig.~\ref{fig:radius}) consistent with 
experiments \cite{Romming:15.1} and micromagnetic models \cite{Bogdanov-1994aa}. 

\begin{figure}[thp]
\centering
\includegraphics[width=0.46\textwidth]{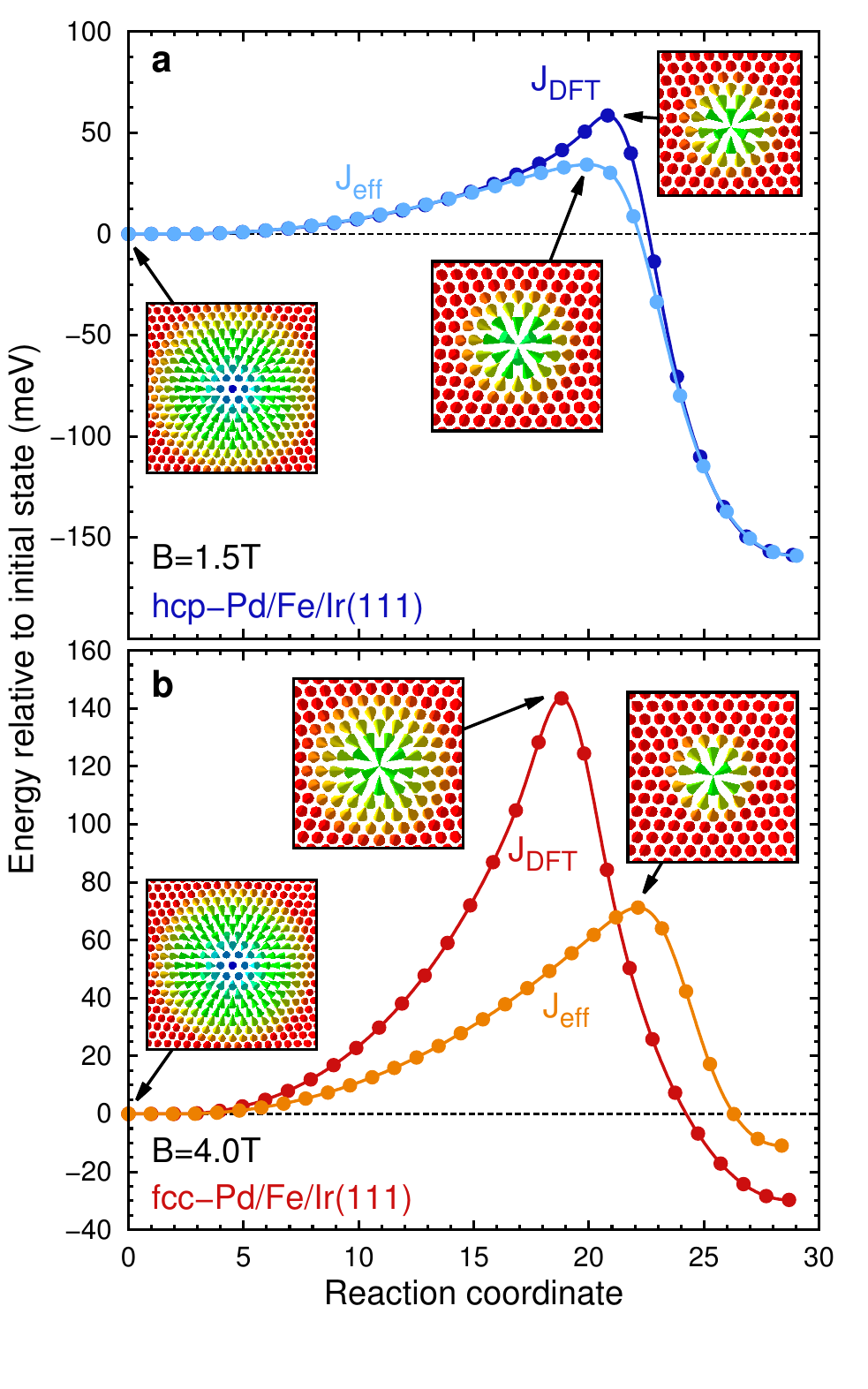}
\caption{
\textbf{Minimum energy paths of skyrmion collapse.}
Energies of the spin configurations during a skyrmion collapse in ({\bf a}) hcp-Pd/Fe/Ir(111) at $B=1.5$~T and ({\bf b}) fcc-Pd/Fe/Ir(111) at $B=4.0$~T 
are shown over the reaction coordinate corresponding to the progress of the collapse. 
The energies are given with respect to the initial state, i.e.~the isolated skyrmion. Simulations using both the effective exchange model (denoted by $J_{\rm eff}$) and taking all exchange constants from
the DFT calculation into account (denoted by $J_{\rm DFT}$) are displayed. Insets show the spin structure of the initial and the saddle point configurations 
for both simulations. 
}
\label{fig:meps}
\end{figure}

For hcp stacking of Pd the radii are nearly the same using either $J_{\rm eff}$ or $J_{\rm DFT}$.
However, isolated skyrmions are metastable for much larger fields if we use the full set of DFT parameters.
For fcc stacking the radii are larger than for hcp stacking and they decrease faster with field considering $J_{\rm eff}$ leading to a crossing of the two curves.
The radii obtained from experiments on Pd/Fe/Ir(111) \cite{Romming:15.1} are very close to those from $J_{\rm DFT}$ for fcc stacking of Pd.
Antiskyrmions can be stabilized only if frustrated exchange is present as in fcc-Pd/Fe/Ir(111). Their radii are smaller than those of skyrmions in the same
system. Below $B=3$~T they become asymmetric, stretched objects and vanish for fields above $B=5.5$~T.

Overall we find a very good agreement between simulations using the full set of exchange constants, the effective parameters and experimental data. 
We conclude that equilibrium properties of skyrmions in Pd/Fe/Ir(111) such as field dependent profiles or zero temperature phase diagrams 
can be accurately described by an effective (or micromagnetic) model for the exchange interactions. 

\textbf{Skyrmion stability.} Now we turn to the collapse of isolated skyrmions into the ferromagnetic state. The minimum energy path for this process
is obtained using the GNEB method (see methods). Fig.~\ref{fig:meps} shows the energy barriers calculated for both Pd stackings at selected field values and both
types of describing the exchange. The initial skyrmion state and the mechanism of collapse are independent of how the exchange interaction is treated. 
The skyrmion shrinks along the path until at the saddle point the inner spins turn into the film plane after which
the topological charge vanishes (see Supplementary Figs.~1 and 2) and the skyrmion can
collapse to the ferromagnetic state (see Supplementary Movies 1 to 4). This mechanism has already been described based on the effective exchange model 
\cite{Bessarab:15.1,Rohart:16.1,Lobanov:16.1,Bessarab:17.1,Rohart:17.1}. However, the 
energy barriers are drastically enhanced if the frustrated exchange is explicitly treated in the simulations (see Supplementary Figs.~1 and 2 
for interaction resolved energy contributions along the reaction path). For the chosen magnetic field values we find an enhancement 
from about 34 to 59 meV for hcp-Pd/Fe/Ir(111) and from 71 to 143 meV for fcc-Pd/Fe/Ir(111). 

\begin{figure}[thp]
\centering
\includegraphics[width=0.4\textwidth,bb=0 0 220 240,clip]{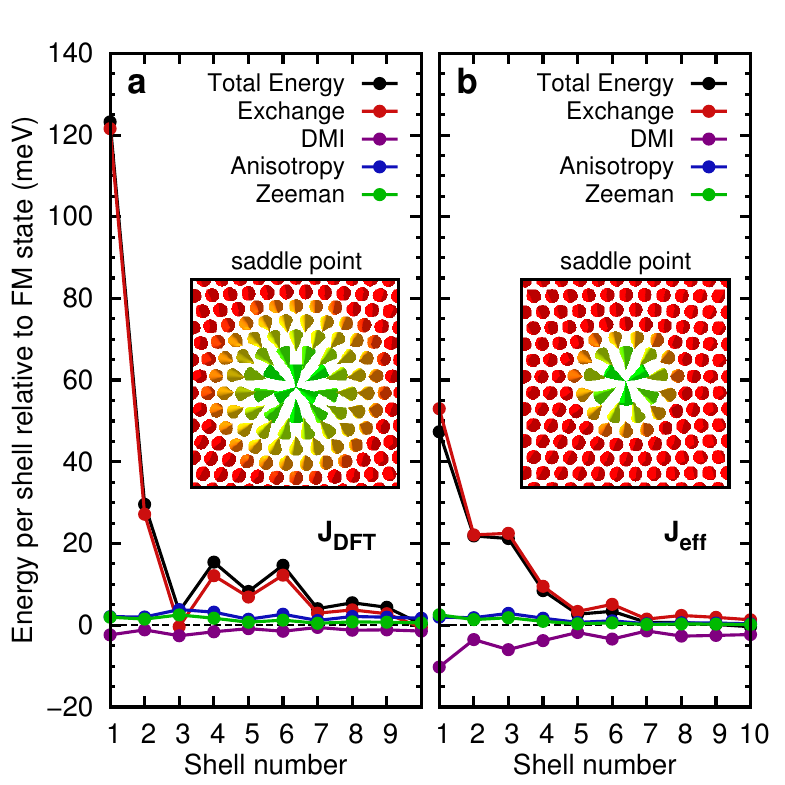}
\caption{
\textbf{Energy contributions at the saddle point.}
Total energy per shell and contributions of the individual interactions are shown relative to the FM state for fcc-Pd/Fe/Ir(111) over the lattice shells of the saddle point for 
(\textbf{a}) DFT ($\textrm{J}_{\textrm{DFT}}$) and (\textbf{b}) effective ($\textrm{J}_{\textrm{eff}}$) parameters. Atoms with the same distance to the midpoint of the saddle point 
are defined as one shell. The insets show the corresponding saddle point configurations.
}
\label{fig:transition-state}
\end{figure}

The spin structure at the saddle point is more extended if the full set of exchange constants is taken into account for
fcc-Pd/Fe/Ir(111) as shown in the insets of Fig.~\ref{fig:transition-state}.
The main contribution to the energy at the saddle point with respect to the FM state 
stems from the exchange interactions while all other interactions are small.
The inner shells of the spin structure give the largest contribution since the angles between adjacent spins are largest. 
This energy is much enhanced if the full set of exchange constants is considered since using an effective exchange $J_{\rm eff}$ underestimates this term. 
The role of the frustrated exchange is apparent also from its energy contribution along the reaction path (see supplementary Fig.~S1)
which exhibits a large barrier -- an effect which is absent for an effective exchange.
For hcp stacking of the Pd layer a similar picture is obtained (see Supplementary Figs.~2 and 3).

\begin{figure}[thp]
\centering
\includegraphics[width=0.46\textwidth]{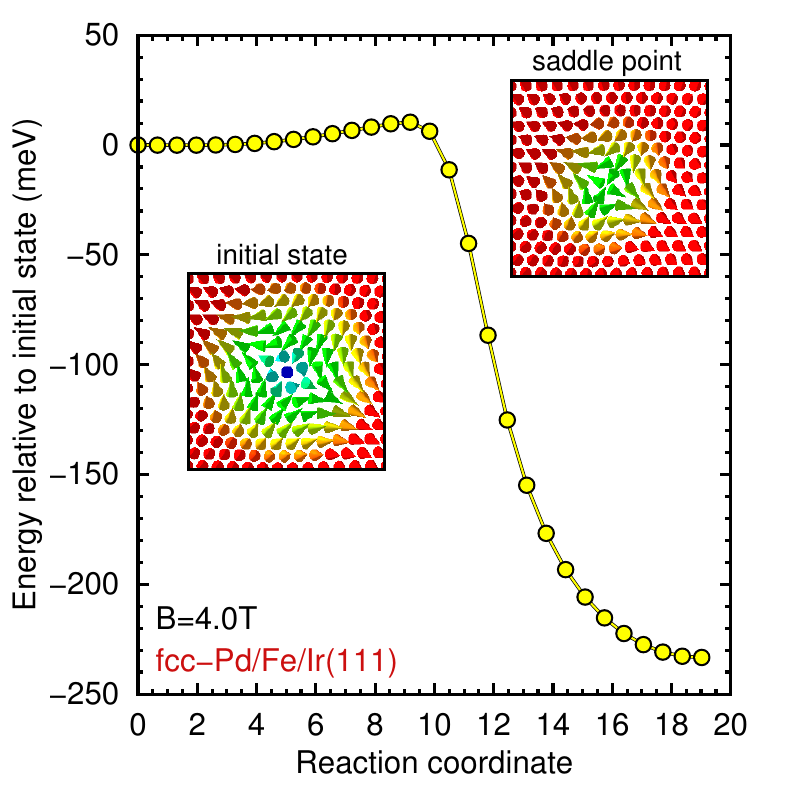}
\caption{
\textbf{Minimum energy path of antiskyrmion collapse.}
Total energies of the spin configurations of an antiskyrmion collapse into the ferromagnetic state over the reaction coordinate corresponding to the progress of the collapse. The energies are given
relative to the initial state, i.e.~the isolated antiskyrmion. Insets show the isolated antiskyrmion and the saddle point configuration. In this simulation the full set of DFT parameters 
for fcc-Pd/Fe/Ir(111) were used.
}
\label{fig:antiskyrmion}
\end{figure}

Frustrated exchange interactions allow the appearance of metastable antiskyrmions \cite{Okubo:12.1,Leonov:15.1,Dupe:16.2,Rosza:17.1}. We find non-vanishing energy barriers only in our simulations
of fcc-Pd/Fe/Ir(111) using the full set of DFT parameters. In this case the exchange interactions alone already lead to a shallow energy minimum in the dispersion
of spin spirals. The mechanism of antiskyrmion collapse is similar to that for skyrmions with a shrinking of diameter and a saddle point configuration with inner spins pointing in the film plane (see insets of Fig.~\ref{fig:antiskyrmion} and Supplementary Movie 5). We obtain a small energy barrier of about 15~meV at a magnetic field of $B=4$~T as shown in Fig.~\ref{fig:antiskyrmion} (see Supplementary 
Fig.~4 for interaction resolved energy contributions along the path). 

\noindent \textbf{Discussion} 

\begin{figure}[thp]
\centering
\includegraphics[width=0.46\textwidth]{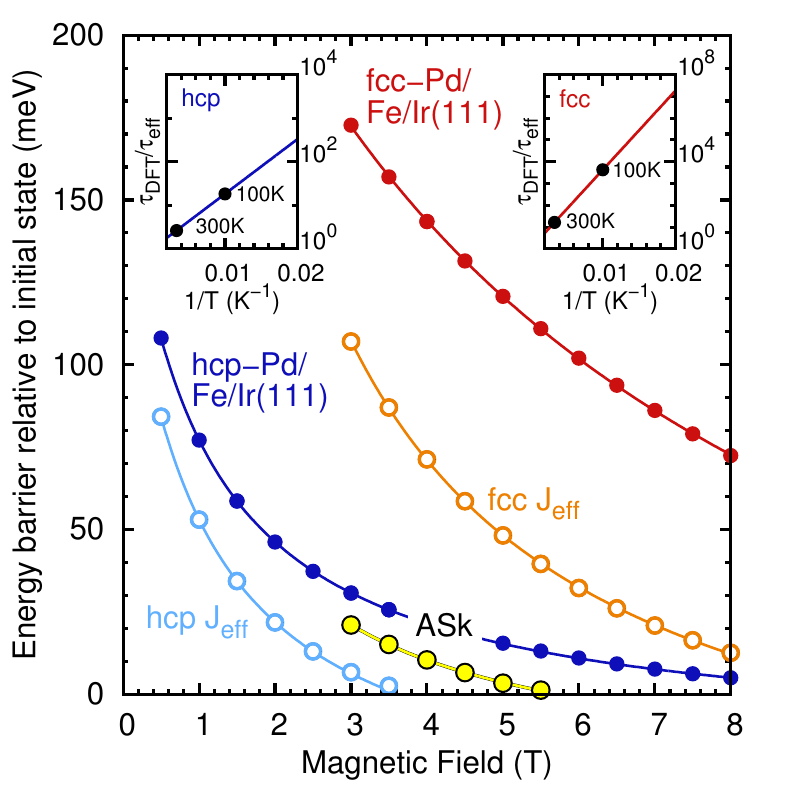}
\caption{
\textbf{Energy barriers of skyrmion collapse vs.~magnetic field and skyrmion lifetime enhancement.}
The energy barriers of isolated skyrmion collapse are shown as a function of applied magnetic field
for the full set of DFT parameters as well as for the effective NN exchange model. 
The energy barriers of isolaed antiskyrmions which are metastable in fcc-Pd/Fe/Ir(111) using the DFT parameters
are also displayed. The energy barriers are defined with respect to the initial states of isolated skyrmions or antiskyrmions.
The inset shows the temperature dependence of the skyrmion lifetime enhancement $\tau_{\rm DFT}/\tau_{\rm eff}$ obtained 
from the energy barrier difference between DFT and effective parameters. Since the energy difference varies only by a few meV 
with field we take the values 72~meV for fcc ($B=4$~T) and 25~meV for hcp ($B=1.5$~T) stacking of Pd.
}
\label{fig:barrier_over_b}
\end{figure}

\noindent  
The energy barriers for skyrmion collapse exhibit a nonlinear dependence on the external magnetic field as shown in Fig.~\ref{fig:barrier_over_b}.  
For fcc-Pd/Fe/Ir(111) the barriers above the critical field of $B^{\rm eff}_{\rm c}=3.6$~T are of the same order if we use $J_{\rm eff}$ 
as given by Hagemeister {\it et al.} \cite{Hagemeister:15.1} based on Monte-Carlo simulations with an effective NN exchange. 
However, for all field values the barriers are drastically enhanced if the frustrated exchange in this system is treated
properly. As a consequence isolated skyrmions are metastable up to much higher magnetic fields.
Furthermore, antiskyrmions are only metastable if the full exchange interaction is taken into account with barriers
on the order of 10 to 20~meV.

Even for hcp-Pd/Fe/Ir(111) which shows a ferromagnetic ground state at all field values there is a significant difference between the barriers 
for $J_{\rm eff}$ vs.~$J_{\rm DFT}$. Despite the very similar zero temperature phase diagram (cf.~Fig.~\ref{fig:dispersion_stability}b,c)
isolated skyrmions are metastable at much larger fields for $J_{\rm DFT}$.

Within transition-state theory \cite{Bessarab:12.1} the lifetime $\tau$ of skyrmions depends exponentially on the energy barrier 
$\Delta E$, i.e.~$\tau = \tau_0 \exp{(\Delta E/{k_{\rm B} T})}$ where $T$ is the temperature and $1/\tau_0$ is the attempt frequency. 
The enhancement of the lifetime due to taking exchange beyond NN into account for the barrier is 
$\tau_{\rm DFT}/\tau_{\rm eff}=\exp{([\Delta E_{\rm DFT}-\Delta E_{\rm eff}]/{k_{\rm B}T})}$.
For both hcp and fcc stacking of Pd there is a huge enhancement by one or three orders of magnitude at 100~K, respectively (see insets of Fig.~\ref{fig:barrier_over_b}). 
Even at room temperature there is a large increase of skyrmion lifetime.

The key observation is that at the saddle point which defines the height of the energy barrier the angles between
adjacent spins are large and the energy of this configuration cannot be accurately described here by the effective NN 
exchange or micromagnetic model. 
From the energy dispersion of spin spirals (cf.~Fig.~\ref{fig:dispersion_stability}a) this is obvious since the
effective NN exchange can only describe $E(q)$ well close to the ferromagnetic state for small canting angles
between adjacent spins while it breaks down at large $q$ corresponding to large NN angles. 

The effective exchange model even with parameters obtained from experiment \cite{Romming:15.1,Hagemeister:15.1,Leonov:16.1}
is limited to describe skyrmion properties such as skyrmion profiles or zero temperature phase diagrams which rely on states
close to the energy minimum. For all properties which require the treatment
of configurations with large angles between adjacent spins such as skyrmion stability it is not sufficient.
Therefore, properly treating exchange is also essential to simulate thermal fluctuations in order to obtain phase diagrams 
at finite temperature and phase transition temperatures \cite{Rosza:16.1}. 

Our work shows the importance of accounting 
for the neighbor resolved exchange interaction even in systems with a ferromagnetic ground state
and that engineering exchange frustration \cite{Dupe:16.1} is a promising route towards enhanced skyrmion stability.
Although we focus here on a particular system, the conclusions we arrive at are rather general and should be valid in 
other itinerant electron magnets where frustrated, long-range exchange is a typical feature.

\noindent \textbf{Methods}

\noindent {\bf First-principles calculations.} We have explored the ultrathin film system Pd/Fe/Ir(111) from first-principles based on the
full-potential linearized augmented plane wave method as implemented in the FLEUR code (www.flapw.de). Within 
this approach we can calculate the total energy 
of non-collinear magnetic structures such as spin spirals~\cite{Kurz-SS} including 
the DMI in first order perturbation theory with respect to the spin-orbit coupling~\cite{Heide-DMI}. 
We have used a two-dimensional hexagonal p$(1\times1)$ unit cell within each layer and 
the in plane-lattice parameter of the Ir(111) surface as obtained from DFT in
Ref.~[\onlinecite{Dupe:14.1}]. The relaxed interlayer distances were also taken
from Ref.~[\onlinecite{Dupe:14.1}].
The magnetic properties were obtained within the local density approximation~\cite{doi-10.1139/p80-159}
as previously \cite{Ferriani2008,Heinze:11.1,Dupe:14.1}. 
All calculations have been carried out with a plane wave cutoff of $k_{\mathrm{max}}=4.3$~a.u.$^{-1}$. The muffin tin radii were set to $2.23$~a.u.~for Fe 
and to $2.31$~a.u.~for Pd and Ir. We have used a $44 \times 44$~k-point mesh in the full two-dimensional Brillouin zone.
In order to obtain quantitative parameters for the exchange interactions, we have converged the small energy minimum of the dispersion curve
(Fig.~1(a)) with respect to the number of Ir substrate layers. 
With 5 layers of Ir substrate \cite{Dupe:14.1} the energy minimum is $E_{\mathrm{min}}=-1.1$~meV/Fe (without spin-orbit coupling). 
Here, we have used asymmetric films with one layer Pd, one layer Fe and
15 layers of Ir substrate which leads to an energy minimum of $E_{\mathrm{min}}=-0.06$~meV/Fe. 
To determine the exchange constants, we have used the energy dispersion curve of spin spirals $E(q)$ for films with 15 Ir substrate layers
for the spiral vector $q$ in $[0,1/4]$ and with 5 Ir layers for $q$ in $[1/4,2/3]$ (in units of $2\pi/a$) along the $\bar{\Gamma} \bar{\rm K}$ direction. 
The DM interaction has been taken from the calculations with 5 Ir substrate layers \cite{Dupe:14.1} as it did not change with substrate thickness.

\noindent {\bf Spin dynamics simulations.} 
In order to calculate the energy differences between the FM, SkX and SS phases for the low temperature phase diagrams and to relax the isolated skyrmions and antiskyrmions,
we used the Landau-Lifshitz equation of spin dynamics:
\begin{equation}
\hbar \frac{d \mathbf m_i}{dt} = \frac{\partial H}{\partial \mathbf m_i} \times \mathbf m_i - \alpha \left( \frac{\partial H}{\partial \mathbf m_i} \times \mathbf m_i \right)  \times \mathbf m_i
\end{equation}
where $\alpha$ is the damping parameter, $\mathbf m_i$ is a single spin and $H$ is the Hamiltonian given in Eq.~(1).
We have used damping parameters from $\alpha=0.5$ to $\alpha=0.05$ and time steps ranging from 0.05 fs to  0.1 fs to carefully relax the structures. 
The simulations are carried out on a timescale of 100 to 500 pico seconds and the equation of motion were solved with the semi-implicit integrator as proposed by Mentink \textit{et al.} \cite{Mentink2010}

\noindent {\bf Geodesic nudged elastic band method.}
Minimum energy paths (MEPs) for skyrmion and antiskyrmion annihilation processes were identified using the geodesic nudged elastic band (GNEB) method [24]. The GNEB method involves taking some initially generated path between the energy minima, and systematically bringing that to the MEP. A path is represented by a discrete chain of states, or 'images', of the system, where the first one corresponds to the skyrmion or antiskyrmion configuration while the last one corresponds to the ferromagnetic state. At each image, effective field is calculated and its component along a local tangent to the path is substituted by an artificial spring force between the images which ensures uniform distribution of the images along the path. This modified effective field is substituted in Eq.~(2) 
in the overdamped regime 
and the whole chain of images evolved till convergence. The final, relaxed position of images gives discrete representation of the MEP. The energy maximum along the MEP corresponds to the saddle point (SP) on the multidimensional energy surface of the system and defines the energy barrier separating the stable states. In order to determine the maximum energy accurately, the highest energy image can be treated separately during the iterative optimization and made to move uphill in energy along the path. The effective field on this climbing image (CI) is calculated by deactivating the spring force acting on it and inverting the parallel component of the gradient field. After the CI-GNEB calculation has converged, the position of the CI coincides with the SP along the MEP and gives an accurate value of the SP energy.

\vspace*{-0.75cm}
\section*{Acknowledgments}
We gratefully acknowledge computing time at the supercomputer of the North-German Supercomputing Alliance (HLRN).
This project has received funding from the European Union’s Horizon 2020 research and innovation programme 
under grant agreement No 665095 (FET-Open project MAGicSky).
P.F.B.~acknowledges support from the Icelandic Research Fund (Grant No.~163048-052) and the mega-grant of the Ministry of Education and Science of the Russian Federation (grant no.~14.Y26.31.0015). B.D.~thanks the Deutsche Forschungsgemeinschaft for funding within the project DU1489/2-1. 
A.D.~acknowledges financial support from Swedish e-science Research Centre (SeRC), Vetenskapsr{\aa}det (grant numbers VR 2015-04608 and VR 2016-05980), and Swedish Energy Agency (grant number STEM P40147-1). Some of the computations were performed on resources provided by the Swedish National Infrastructure for Computing (SNIC) at the National Supercomputer Center (NSC), Link\"oping University, the PDC Centre for High Performance Computing (PDC-HPC), KTH, and the High Performance Computing Center North (HPC2N), Ume{\aa} University.

\vspace*{-0.5cm}
\section*{Author contributions}
S.v.M. performed the spin dynamics simulations. S.v.M. and P.F.B. performed the GNEB calculations. 
B.D. performed the first-principles calculations. B.D., P.F.B., A.D., and S.H. conceived the project. 
S.v.M. and S.H. wrote the manuscript.
All authors analyzed and discussed the data and contributed to the writing of the paper.

\vspace*{-.95cm}
\section*{Additional information}
Competing financial interests: The authors declare no competing financial interests.


\vspace*{-.5cm}

\begin{thebibliography}{50}
\expandafter\ifx\csname natexlab\endcsname\relax\def\natexlab#1{#1}\fi
\expandafter\ifx\csname bibnamefont\endcsname\relax
  \def\bibnamefont#1{#1}\fi
\expandafter\ifx\csname bibfnamefont\endcsname\relax
  \def\bibfnamefont#1{#1}\fi
\expandafter\ifx\csname citenamefont\endcsname\relax
  \def\citenamefont#1{#1}\fi
\expandafter\ifx\csname url\endcsname\relax
  \def\url#1{\texttt{#1}}\fi
\expandafter\ifx\csname urlprefix\endcsname\relax\def\urlprefix{URL }\fi
\providecommand{\bibinfo}[2]{#2}
\providecommand{\eprint}[2][]{\url{#2}}

\bibitem[{\citenamefont{Bogdanov and Yablonskii}(1989)}]{Bogdanov1989}
\bibinfo{author}{\bibfnamefont{A.}~\bibnamefont{Bogdanov}} \bibnamefont{and}
  \bibinfo{author}{\bibfnamefont{D.~A.} \bibnamefont{Yablonskii}},
  \bibinfo{journal}{Sov. Phys. JETP} \textbf{\bibinfo{volume}{68}},
  \bibinfo{pages}{101} (\bibinfo{year}{1989}).

\bibitem[{\citenamefont{Bogdanov and
  Hubert}(1994{\natexlab{a}})}]{Bogdanov1994}
\bibinfo{author}{\bibfnamefont{A.}~\bibnamefont{Bogdanov}} \bibnamefont{and}
  \bibinfo{author}{\bibfnamefont{A.}~\bibnamefont{Hubert}},
  \bibinfo{journal}{J. Mag. Mag. Mat.} \textbf{\bibinfo{volume}{138}},
  \bibinfo{pages}{255} (\bibinfo{year}{1994}{\natexlab{a}}).

\bibitem[{\citenamefont{M\"uhlbauer et~al.}(2009)\citenamefont{M\"uhlbauer,
  Binz, Jonietz, Pfleiderer, Rosch, Neubauer, Georgii, and
  B\"oni}}]{Muehlbauer:09.1}
\bibinfo{author}{\bibfnamefont{S.}~\bibnamefont{M\"uhlbauer}},
  \bibinfo{author}{\bibfnamefont{B.}~\bibnamefont{Binz}},
  \bibinfo{author}{\bibfnamefont{F.}~\bibnamefont{Jonietz}},
  \bibinfo{author}{\bibfnamefont{C.}~\bibnamefont{Pfleiderer}},
  \bibinfo{author}{\bibfnamefont{A.}~\bibnamefont{Rosch}},
  \bibinfo{author}{\bibfnamefont{A.}~\bibnamefont{Neubauer}},
  \bibinfo{author}{\bibfnamefont{R.}~\bibnamefont{Georgii}}, \bibnamefont{and}
  \bibinfo{author}{\bibfnamefont{P.}~\bibnamefont{B\"oni}},
  \bibinfo{journal}{Science} \textbf{\bibinfo{volume}{323}},
  \bibinfo{pages}{915} (\bibinfo{year}{2009}).

\bibitem[{\citenamefont{Yu et~al.}(2010)\citenamefont{Yu, Onose, Kanazawa,
  Park, Han, Matsui, Nagaosa, and Tokura}}]{Yu:10.1}
\bibinfo{author}{\bibfnamefont{X.~Z.} \bibnamefont{Yu}},
  \bibinfo{author}{\bibfnamefont{Y.}~\bibnamefont{Onose}},
  \bibinfo{author}{\bibfnamefont{N.}~\bibnamefont{Kanazawa}},
  \bibinfo{author}{\bibfnamefont{J.~H.} \bibnamefont{Park}},
  \bibinfo{author}{\bibfnamefont{J.~H.} \bibnamefont{Han}},
  \bibinfo{author}{\bibfnamefont{Y.}~\bibnamefont{Matsui}},
  \bibinfo{author}{\bibfnamefont{N.}~\bibnamefont{Nagaosa}}, \bibnamefont{and}
  \bibinfo{author}{\bibfnamefont{Y.}~\bibnamefont{Tokura}},
  \bibinfo{journal}{Nature} \textbf{\bibinfo{volume}{465}},
  \bibinfo{pages}{901–904} (\bibinfo{year}{2010}).

\bibitem[{\citenamefont{Seki et~al.}(2012)\citenamefont{Seki, Yu, Ishiwata, and
  Tokura}}]{Seki2012}
\bibinfo{author}{\bibfnamefont{S.}~\bibnamefont{Seki}},
  \bibinfo{author}{\bibfnamefont{X.~Z.} \bibnamefont{Yu}},
  \bibinfo{author}{\bibfnamefont{S.}~\bibnamefont{Ishiwata}}, \bibnamefont{and}
  \bibinfo{author}{\bibfnamefont{Y.}~\bibnamefont{Tokura}},
  \bibinfo{journal}{Science} \textbf{\bibinfo{volume}{336}},
  \bibinfo{pages}{198} (\bibinfo{year}{2012}).

\bibitem[{\citenamefont{Moriya}(1960)}]{Moriya}
\bibinfo{author}{\bibfnamefont{T.}~\bibnamefont{Moriya}},
  \bibinfo{journal}{Phys. Rev.} \textbf{\bibinfo{volume}{120}},
  \bibinfo{pages}{91} (\bibinfo{year}{1960}).

\bibitem[{\citenamefont{Dzyaloshinskii}(1957)}]{Dzyaloshinskii}
\bibinfo{author}{\bibfnamefont{I.~E.} \bibnamefont{Dzyaloshinskii}},
  \bibinfo{journal}{Sov. Phys. JETP} \textbf{\bibinfo{volume}{5}},
  \bibinfo{pages}{1259} (\bibinfo{year}{1957}).

\bibitem[{\citenamefont{Cr\'epieux and Lacroix}(1998)}]{Crepieux1998}
\bibinfo{author}{\bibfnamefont{A.}~\bibnamefont{Cr\'epieux}} \bibnamefont{and}
  \bibinfo{author}{\bibfnamefont{C.}~\bibnamefont{Lacroix}},
  \bibinfo{journal}{J. Mag. Mag. Mat.} \textbf{\bibinfo{volume}{182}},
  \bibinfo{pages}{341} (\bibinfo{year}{1998}).

\bibitem[{\citenamefont{Bode et~al.}(2007)\citenamefont{Bode, Heide, von
  Bergmann, Ferriani, Heinze, Bihlmayer, Kubetzka, Pietzsch, Bl\"ugel, and
  Wiesendanger}}]{Bode2007}
\bibinfo{author}{\bibfnamefont{M.}~\bibnamefont{Bode}},
  \bibinfo{author}{\bibfnamefont{M.}~\bibnamefont{Heide}},
  \bibinfo{author}{\bibfnamefont{K.}~\bibnamefont{von Bergmann}},
  \bibinfo{author}{\bibfnamefont{P.}~\bibnamefont{Ferriani}},
  \bibinfo{author}{\bibfnamefont{S.}~\bibnamefont{Heinze}},
  \bibinfo{author}{\bibfnamefont{G.}~\bibnamefont{Bihlmayer}},
  \bibinfo{author}{\bibfnamefont{A.}~\bibnamefont{Kubetzka}},
  \bibinfo{author}{\bibfnamefont{O.}~\bibnamefont{Pietzsch}},
  \bibinfo{author}{\bibfnamefont{S.}~\bibnamefont{Bl\"ugel}}, \bibnamefont{and}
  \bibinfo{author}{\bibfnamefont{R.}~\bibnamefont{Wiesendanger}},
  \bibinfo{journal}{Nature} \textbf{\bibinfo{volume}{447}},
  \bibinfo{pages}{190} (\bibinfo{year}{2007}).

\bibitem[{\citenamefont{Ferriani et~al.}(2008)\citenamefont{Ferriani, von
  Bergmann, Vedmedenko, Heinze, Bode, Heide, Bihlmayer, Bl\"ugel, and
  Wiesendanger}}]{Ferriani2008}
\bibinfo{author}{\bibfnamefont{P.}~\bibnamefont{Ferriani}},
  \bibinfo{author}{\bibfnamefont{K.}~\bibnamefont{von Bergmann}},
  \bibinfo{author}{\bibfnamefont{E.~Y.} \bibnamefont{Vedmedenko}},
  \bibinfo{author}{\bibfnamefont{S.}~\bibnamefont{Heinze}},
  \bibinfo{author}{\bibfnamefont{M.}~\bibnamefont{Bode}},
  \bibinfo{author}{\bibfnamefont{M.}~\bibnamefont{Heide}},
  \bibinfo{author}{\bibfnamefont{G.}~\bibnamefont{Bihlmayer}},
  \bibinfo{author}{\bibfnamefont{S.}~\bibnamefont{Bl\"ugel}}, \bibnamefont{and}
  \bibinfo{author}{\bibfnamefont{R.}~\bibnamefont{Wiesendanger}},
  \bibinfo{journal}{Phys. Rev. Lett.} \textbf{\bibinfo{volume}{101}},
  \bibinfo{pages}{027201} (\bibinfo{year}{2008}).

\bibitem[{\citenamefont{Heide et~al.}(2008)\citenamefont{Heide, Bihlmayer, and
  Bl\"ugel}}]{Heide2008}
\bibinfo{author}{\bibfnamefont{M.}~\bibnamefont{Heide}},
  \bibinfo{author}{\bibfnamefont{G.}~\bibnamefont{Bihlmayer}},
  \bibnamefont{and} \bibinfo{author}{\bibfnamefont{S.}~\bibnamefont{Bl\"ugel}},
  \bibinfo{journal}{Phys. Rev. B} \textbf{\bibinfo{volume}{78}},
  \bibinfo{pages}{140403} (\bibinfo{year}{2008}).

\bibitem[{\citenamefont{Meckler et~al.}(2009)\citenamefont{Meckler, Mikuszeit,
  Pre{\ss}ler, Vedmedenko, Pietzsch, and Wiesendanger}}]{Meckler2009}
\bibinfo{author}{\bibfnamefont{S.}~\bibnamefont{Meckler}},
  \bibinfo{author}{\bibfnamefont{N.}~\bibnamefont{Mikuszeit}},
  \bibinfo{author}{\bibfnamefont{A.}~\bibnamefont{Pre{\ss}ler}},
  \bibinfo{author}{\bibfnamefont{E.~Y.} \bibnamefont{Vedmedenko}},
  \bibinfo{author}{\bibfnamefont{O.}~\bibnamefont{Pietzsch}}, \bibnamefont{and}
  \bibinfo{author}{\bibfnamefont{R.}~\bibnamefont{Wiesendanger}},
  \bibinfo{journal}{Phys. Rev. Lett.} \textbf{\bibinfo{volume}{103}},
  \bibinfo{pages}{157201} (\bibinfo{year}{2009}).

\bibitem[{\citenamefont{Emori et~al.}(2013)\citenamefont{Emori, Bauer, Ahn,
  Martinez, and Beach}}]{nmat3675}
\bibinfo{author}{\bibfnamefont{S.}~\bibnamefont{Emori}},
  \bibinfo{author}{\bibfnamefont{U.}~\bibnamefont{Bauer}},
  \bibinfo{author}{\bibfnamefont{S.-M.} \bibnamefont{Ahn}},
  \bibinfo{author}{\bibfnamefont{E.}~\bibnamefont{Martinez}}, \bibnamefont{and}
  \bibinfo{author}{\bibfnamefont{G.~S.~D.} \bibnamefont{Beach}},
  \bibinfo{journal}{Nature materials} \textbf{\bibinfo{volume}{12}},
  \bibinfo{pages}{611} (\bibinfo{year}{2013}).

\bibitem[{\citenamefont{Ryu et~al.}(2013)\citenamefont{Ryu, Thomas, Yang, and
  Parkin}}]{Ryu2013}
\bibinfo{author}{\bibfnamefont{K.-S.} \bibnamefont{Ryu}},
  \bibinfo{author}{\bibfnamefont{L.}~\bibnamefont{Thomas}},
  \bibinfo{author}{\bibfnamefont{S.-H.} \bibnamefont{Yang}}, \bibnamefont{and}
  \bibinfo{author}{\bibfnamefont{S.}~\bibnamefont{Parkin}},
  \bibinfo{journal}{Nature Nanotech.} \textbf{\bibinfo{volume}{8}},
  \bibinfo{pages}{527} (\bibinfo{year}{2013}).

\bibitem[{\citenamefont{Thiaville et~al.}(2012)\citenamefont{Thiaville, Rohart,
  Ju{\'e}, Cros, and Fert}}]{thiaville2012dynamics}
\bibinfo{author}{\bibfnamefont{A.}~\bibnamefont{Thiaville}},
  \bibinfo{author}{\bibfnamefont{S.}~\bibnamefont{Rohart}},
  \bibinfo{author}{\bibfnamefont{{\'E}.}~\bibnamefont{Ju{\'e}}},
  \bibinfo{author}{\bibfnamefont{V.}~\bibnamefont{Cros}}, \bibnamefont{and}
  \bibinfo{author}{\bibfnamefont{A.}~\bibnamefont{Fert}}, \bibinfo{journal}{EPL
  (Europhysics Letters)} \textbf{\bibinfo{volume}{100}}, \bibinfo{pages}{57002}
  (\bibinfo{year}{2012}).

\bibitem[{\citenamefont{Heinze et~al.}(2011)\citenamefont{Heinze, v.~Bergmann,
  Menzel, Brede, Kubetzka, Wiesendanger, Bihlmayer, and
  Bl\"ugel}}]{Heinze:11.1}
\bibinfo{author}{\bibfnamefont{S.}~\bibnamefont{Heinze}},
  \bibinfo{author}{\bibfnamefont{K.}~\bibnamefont{v.~Bergmann}},
  \bibinfo{author}{\bibfnamefont{M.}~\bibnamefont{Menzel}},
  \bibinfo{author}{\bibfnamefont{J.}~\bibnamefont{Brede}},
  \bibinfo{author}{\bibfnamefont{A.}~\bibnamefont{Kubetzka}},
  \bibinfo{author}{\bibfnamefont{R.}~\bibnamefont{Wiesendanger}},
  \bibinfo{author}{\bibfnamefont{G.}~\bibnamefont{Bihlmayer}},
  \bibnamefont{and} \bibinfo{author}{\bibfnamefont{S.}~\bibnamefont{Bl\"ugel}},
  \bibinfo{journal}{Nature Phys.} \textbf{\bibinfo{volume}{7}},
  \bibinfo{pages}{713} (\bibinfo{year}{2011}).

\bibitem[{\citenamefont{Romming et~al.}(2013)\citenamefont{Romming, Hanneken,
  Menzel, Bickel, Wolter, von Bergmann, Kubetzka, and
  Wiesendanger}}]{Romming:13.1}
\bibinfo{author}{\bibfnamefont{N.}~\bibnamefont{Romming}},
  \bibinfo{author}{\bibfnamefont{C.}~\bibnamefont{Hanneken}},
  \bibinfo{author}{\bibfnamefont{M.}~\bibnamefont{Menzel}},
  \bibinfo{author}{\bibfnamefont{J.~E.} \bibnamefont{Bickel}},
  \bibinfo{author}{\bibfnamefont{B.}~\bibnamefont{Wolter}},
  \bibinfo{author}{\bibfnamefont{K.}~\bibnamefont{von Bergmann}},
  \bibinfo{author}{\bibfnamefont{A.}~\bibnamefont{Kubetzka}}, \bibnamefont{and}
  \bibinfo{author}{\bibfnamefont{R.}~\bibnamefont{Wiesendanger}},
  \bibinfo{journal}{Science} \textbf{\bibinfo{volume}{341}},
  \bibinfo{pages}{636} (\bibinfo{year}{2013}).

\bibitem[{\citenamefont{Baibich et~al.}(1988)\citenamefont{Baibich, Broto,
  Fert, Van~Dau, Petroff, Etienne, Creuzet, Friederich, and
  Chazelas}}]{PhysRevLett.61.2472}
\bibinfo{author}{\bibfnamefont{M.~N.} \bibnamefont{Baibich}},
  \bibinfo{author}{\bibfnamefont{J.~M.} \bibnamefont{Broto}},
  \bibinfo{author}{\bibfnamefont{A.}~\bibnamefont{Fert}},
  \bibinfo{author}{\bibfnamefont{F.~N.} \bibnamefont{Van~Dau}},
  \bibinfo{author}{\bibfnamefont{F.}~\bibnamefont{Petroff}},
  \bibinfo{author}{\bibfnamefont{P.}~\bibnamefont{Etienne}},
  \bibinfo{author}{\bibfnamefont{G.}~\bibnamefont{Creuzet}},
  \bibinfo{author}{\bibfnamefont{A.}~\bibnamefont{Friederich}},
  \bibnamefont{and} \bibinfo{author}{\bibfnamefont{J.}~\bibnamefont{Chazelas}},
  \bibinfo{journal}{Phys. Rev. Lett.} \textbf{\bibinfo{volume}{61}},
  \bibinfo{pages}{2472} (\bibinfo{year}{1988}).

\bibitem[{\citenamefont{Binasch et~al.}(1989)\citenamefont{Binasch, Gr\"unberg,
  Saurenbach, and Zinn}}]{PhysRevB.39.4828}
\bibinfo{author}{\bibfnamefont{G.}~\bibnamefont{Binasch}},
  \bibinfo{author}{\bibfnamefont{P.}~\bibnamefont{Gr\"unberg}},
  \bibinfo{author}{\bibfnamefont{F.}~\bibnamefont{Saurenbach}},
  \bibnamefont{and} \bibinfo{author}{\bibfnamefont{W.}~\bibnamefont{Zinn}},
  \bibinfo{journal}{Phys. Rev. B} \textbf{\bibinfo{volume}{39}},
  \bibinfo{pages}{4828} (\bibinfo{year}{1989}).

\bibitem[{\citenamefont{Fert et~al.}(2013)\citenamefont{Fert, Cros, and
  Sampaio}}]{Fert:13.1}
\bibinfo{author}{\bibfnamefont{A.}~\bibnamefont{Fert}},
  \bibinfo{author}{\bibfnamefont{V.}~\bibnamefont{Cros}}, \bibnamefont{and}
  \bibinfo{author}{\bibfnamefont{J.}~\bibnamefont{Sampaio}},
  \bibinfo{journal}{Nature Nanotech.} \textbf{\bibinfo{volume}{8}},
  \bibinfo{pages}{152} (\bibinfo{year}{2013}).

\bibitem[{\citenamefont{Sampaio et~al.}(2013)\citenamefont{Sampaio, Cros,
  Rohart, Thiaville, and Fert}}]{Sampaio:13.1}
\bibinfo{author}{\bibfnamefont{J.}~\bibnamefont{Sampaio}},
  \bibinfo{author}{\bibfnamefont{V.}~\bibnamefont{Cros}},
  \bibinfo{author}{\bibfnamefont{S.}~\bibnamefont{Rohart}},
  \bibinfo{author}{\bibfnamefont{A.}~\bibnamefont{Thiaville}},
  \bibnamefont{and} \bibinfo{author}{\bibfnamefont{A.}~\bibnamefont{Fert}},
  \bibinfo{journal}{Nature Nanotech.} \textbf{\bibinfo{volume}{8}},
  \bibinfo{pages}{839–844} (\bibinfo{year}{2013}).

\bibitem[{\citenamefont{Kiselev et~al.}(2011)\citenamefont{Kiselev, Bogdanov,
  Sch\"afer, and R\"o{\ss}ler}}]{Kiselev2011}
\bibinfo{author}{\bibfnamefont{N.}~\bibnamefont{Kiselev}},
  \bibinfo{author}{\bibfnamefont{A.~N.} \bibnamefont{Bogdanov}},
  \bibinfo{author}{\bibfnamefont{R.}~\bibnamefont{Sch\"afer}},
  \bibnamefont{and} \bibinfo{author}{\bibfnamefont{U.~K.}
  \bibnamefont{R\"o{\ss}ler}}, \bibinfo{journal}{J. Phys. D}
  \textbf{\bibinfo{volume}{44}}, \bibinfo{pages}{392001}
  (\bibinfo{year}{2011}).

\bibitem[{\citenamefont{Nagaosa and Tokura}(2013)}]{Nagaosa:13.1}
\bibinfo{author}{\bibfnamefont{N.}~\bibnamefont{Nagaosa}} \bibnamefont{and}
  \bibinfo{author}{\bibfnamefont{Y.}~\bibnamefont{Tokura}},
  \bibinfo{journal}{Nature Nanotech.} \textbf{\bibinfo{volume}{8}},
  \bibinfo{pages}{899} (\bibinfo{year}{2013}).

\bibitem[{\citenamefont{Romming et~al.}(2015)\citenamefont{Romming, Kubetzka,
  Hanneken, von Bergmann, and Wiesendanger}}]{Romming:15.1}
\bibinfo{author}{\bibfnamefont{N.}~\bibnamefont{Romming}},
  \bibinfo{author}{\bibfnamefont{A.}~\bibnamefont{Kubetzka}},
  \bibinfo{author}{\bibfnamefont{C.}~\bibnamefont{Hanneken}},
  \bibinfo{author}{\bibfnamefont{K.}~\bibnamefont{von Bergmann}},
  \bibnamefont{and}
  \bibinfo{author}{\bibfnamefont{R.}~\bibnamefont{Wiesendanger}},
  \bibinfo{journal}{Phys. Rev. Lett.} \textbf{\bibinfo{volume}{114}},
  \bibinfo{pages}{177203} (\bibinfo{year}{2015}).

\bibitem[{\citenamefont{Hanneken et~al.}(2015)\citenamefont{Hanneken, Otte,
  Kubetzka, Dup\'e, Romming, von Bergmann, Wiesendanger, and
  Heinze}}]{Hanneken:15.1}
\bibinfo{author}{\bibfnamefont{C.}~\bibnamefont{Hanneken}},
  \bibinfo{author}{\bibfnamefont{F.}~\bibnamefont{Otte}},
  \bibinfo{author}{\bibfnamefont{A.}~\bibnamefont{Kubetzka}},
  \bibinfo{author}{\bibfnamefont{B.}~\bibnamefont{Dup\'e}},
  \bibinfo{author}{\bibfnamefont{N.}~\bibnamefont{Romming}},
  \bibinfo{author}{\bibfnamefont{K.}~\bibnamefont{von Bergmann}},
  \bibinfo{author}{\bibfnamefont{R.}~\bibnamefont{Wiesendanger}},
  \bibnamefont{and} \bibinfo{author}{\bibfnamefont{S.}~\bibnamefont{Heinze}},
  \bibinfo{journal}{Nature Nanotech.} \textbf{\bibinfo{volume}{10}},
  \bibinfo{pages}{1039} (\bibinfo{year}{2015}).

\bibitem[{\citenamefont{Hagemeister et~al.}(2015)\citenamefont{Hagemeister,
  Romming, von Bergmann, Vedmedenko, and Wiesendanger}}]{Hagemeister:15.1}
\bibinfo{author}{\bibfnamefont{J.}~\bibnamefont{Hagemeister}},
  \bibinfo{author}{\bibfnamefont{N.}~\bibnamefont{Romming}},
  \bibinfo{author}{\bibfnamefont{K.}~\bibnamefont{von Bergmann}},
  \bibinfo{author}{\bibfnamefont{E.~Y.} \bibnamefont{Vedmedenko}},
  \bibnamefont{and}
  \bibinfo{author}{\bibfnamefont{R.}~\bibnamefont{Wiesendanger}},
  \bibinfo{journal}{Nature Comm.} \textbf{\bibinfo{volume}{6}},
  \bibinfo{pages}{8455} (\bibinfo{year}{2015}).

\bibitem[{\citenamefont{Leonov et~al.}(2016)\citenamefont{Leonov, Monchesky,
  Romming, Kubetzka, Bogdanov, and Wiesendanger}}]{Leonov:16.1}
\bibinfo{author}{\bibfnamefont{A.~O.} \bibnamefont{Leonov}},
  \bibinfo{author}{\bibfnamefont{T.~L.} \bibnamefont{Monchesky}},
  \bibinfo{author}{\bibfnamefont{N.}~\bibnamefont{Romming}},
  \bibinfo{author}{\bibfnamefont{A.}~\bibnamefont{Kubetzka}},
  \bibinfo{author}{\bibfnamefont{A.~N.} \bibnamefont{Bogdanov}},
  \bibnamefont{and}
  \bibinfo{author}{\bibfnamefont{R.}~\bibnamefont{Wiesendanger}},
  \bibinfo{journal}{New J. Phys.} \textbf{\bibinfo{volume}{18}},
  \bibinfo{pages}{065003} (\bibinfo{year}{2016}).

\bibitem[{\citenamefont{Bessarab et~al.}(2015)\citenamefont{Bessarab, Uzdin,
  and J\'onsson}}]{Bessarab:15.1}
\bibinfo{author}{\bibfnamefont{P.~F.} \bibnamefont{Bessarab}},
  \bibinfo{author}{\bibfnamefont{V.~M.} \bibnamefont{Uzdin}}, \bibnamefont{and}
  \bibinfo{author}{\bibfnamefont{H.}~\bibnamefont{J\'onsson}},
  \bibinfo{journal}{Comp. Phys. Comm.} \textbf{\bibinfo{volume}{196}},
  \bibinfo{pages}{335} (\bibinfo{year}{2015}).

\bibitem[{\citenamefont{Rohart et~al.}(2016)\citenamefont{Rohart, Miltat, and
  Thiaville}}]{Rohart:16.1}
\bibinfo{author}{\bibfnamefont{S.}~\bibnamefont{Rohart}},
  \bibinfo{author}{\bibfnamefont{J.}~\bibnamefont{Miltat}}, \bibnamefont{and}
  \bibinfo{author}{\bibfnamefont{A.}~\bibnamefont{Thiaville}},
  \bibinfo{journal}{Phys. Rev. B} \textbf{\bibinfo{volume}{93}},
  \bibinfo{pages}{214412} (\bibinfo{year}{2016}).

\bibitem[{\citenamefont{Lobanov et~al.}(2016)\citenamefont{Lobanov, J\'onsson,
  and Uzdin}}]{Lobanov:16.1}
\bibinfo{author}{\bibfnamefont{I.~S.} \bibnamefont{Lobanov}},
  \bibinfo{author}{\bibfnamefont{H.}~\bibnamefont{J\'onsson}},
  \bibnamefont{and} \bibinfo{author}{\bibfnamefont{V.~M.} \bibnamefont{Uzdin}},
  \bibinfo{journal}{Phys. Rev. B} \textbf{\bibinfo{volume}{94}},
  \bibinfo{pages}{174418} (\bibinfo{year}{2016}).

\bibitem[{\citenamefont{Uzdin et~al.}(2017)\citenamefont{Uzdin, Potkina,
  Lobanov, Bessarab, and J\'onsson}}]{Uzdin:17.1}
\bibinfo{author}{\bibfnamefont{V.~M.} \bibnamefont{Uzdin}},
  \bibinfo{author}{\bibfnamefont{M.~N.} \bibnamefont{Potkina}},
  \bibinfo{author}{\bibfnamefont{I.~S.} \bibnamefont{Lobanov}},
  \bibinfo{author}{\bibfnamefont{P.~F.} \bibnamefont{Bessarab}},
  \bibnamefont{and}
  \bibinfo{author}{\bibfnamefont{H.}~\bibnamefont{J\'onsson}},
  \bibinfo{journal}{arXiv 1705.02930}  (\bibinfo{year}{2017}).

\bibitem[{\citenamefont{Bessarab}(2017)}]{Bessarab:17.1}
\bibinfo{author}{\bibfnamefont{P.~F.} \bibnamefont{Bessarab}},
  \bibinfo{journal}{Phys. Rev. B} \textbf{\bibinfo{volume}{95}},
  \bibinfo{pages}{136401} (\bibinfo{year}{2017}).

\bibitem[{\citenamefont{Rohart et~al.}(2017)\citenamefont{Rohart, Miltat, and
  Thiaville}}]{Rohart:17.1}
\bibinfo{author}{\bibfnamefont{S.}~\bibnamefont{Rohart}},
  \bibinfo{author}{\bibfnamefont{J.}~\bibnamefont{Miltat}}, \bibnamefont{and}
  \bibinfo{author}{\bibfnamefont{A.}~\bibnamefont{Thiaville}},
  \bibinfo{journal}{Phys. Rev. B} \textbf{\bibinfo{volume}{95}},
  \bibinfo{pages}{136402} (\bibinfo{year}{2017}).

\bibitem[{\citenamefont{Dup\'e et~al.}(2014)\citenamefont{Dup\'e, Hoffmann,
  Paillard, and Heinze}}]{Dupe:14.1}
\bibinfo{author}{\bibfnamefont{B.}~\bibnamefont{Dup\'e}},
  \bibinfo{author}{\bibfnamefont{M.}~\bibnamefont{Hoffmann}},
  \bibinfo{author}{\bibfnamefont{C.}~\bibnamefont{Paillard}}, \bibnamefont{and}
  \bibinfo{author}{\bibfnamefont{S.}~\bibnamefont{Heinze}},
  \bibinfo{journal}{Nature Comm.} \textbf{\bibinfo{volume}{5}},
  \bibinfo{pages}{4030} (\bibinfo{year}{2014}).

\bibitem[{\citenamefont{Simon et~al.}(2014)\citenamefont{Simon, Palot\'as,
  R\'ozsa, Udvardi, and Szunyogh}}]{Simon:14.1}
\bibinfo{author}{\bibfnamefont{E.}~\bibnamefont{Simon}},
  \bibinfo{author}{\bibfnamefont{K.}~\bibnamefont{Palot\'as}},
  \bibinfo{author}{\bibfnamefont{L.}~\bibnamefont{R\'ozsa}},
  \bibinfo{author}{\bibfnamefont{L.}~\bibnamefont{Udvardi}}, \bibnamefont{and}
  \bibinfo{author}{\bibfnamefont{L.}~\bibnamefont{Szunyogh}},
  \bibinfo{journal}{Phys. Rev. B} \textbf{\bibinfo{volume}{90}},
  \bibinfo{pages}{094410} (\bibinfo{year}{2014}).

\bibitem[{\citenamefont{Kurz et~al.}(2004)\citenamefont{Kurz, F\"orster,
  Nordstr\"om, Bihlmayer, and Bl\"ugel}}]{Kurz-SS}
\bibinfo{author}{\bibfnamefont{P.}~\bibnamefont{Kurz}},
  \bibinfo{author}{\bibfnamefont{F.}~\bibnamefont{F\"orster}},
  \bibinfo{author}{\bibfnamefont{L.}~\bibnamefont{Nordstr\"om}},
  \bibinfo{author}{\bibfnamefont{G.}~\bibnamefont{Bihlmayer}},
  \bibnamefont{and} \bibinfo{author}{\bibfnamefont{S.}~\bibnamefont{Bl\"ugel}},
  \bibinfo{journal}{Phys. Rev. B} \textbf{\bibinfo{volume}{69}},
  \bibinfo{pages}{024415} (\bibinfo{year}{2004}).

\bibitem[{\citenamefont{Heide et~al.}(2009)\citenamefont{Heide, Bihlmayer, and
  Bl\"ugel}}]{Heide-DMI}
\bibinfo{author}{\bibfnamefont{M.}~\bibnamefont{Heide}},
  \bibinfo{author}{\bibfnamefont{G.}~\bibnamefont{Bihlmayer}},
  \bibnamefont{and} \bibinfo{author}{\bibfnamefont{S.}~\bibnamefont{Bl\"ugel}},
  \bibinfo{journal}{Physica B: Condensed Matter}
  \textbf{\bibinfo{volume}{404}}, \bibinfo{pages}{2678 }
  (\bibinfo{year}{2009}).

\bibitem[{\citenamefont{Zimmermann et~al.}(2014)\citenamefont{Zimmermann,
  Heide, Bihlmayer, and Bl\"ugel}}]{Zimmermann2014}
\bibinfo{author}{\bibfnamefont{B.}~\bibnamefont{Zimmermann}},
  \bibinfo{author}{\bibfnamefont{M.}~\bibnamefont{Heide}},
  \bibinfo{author}{\bibfnamefont{G.}~\bibnamefont{Bihlmayer}},
  \bibnamefont{and} \bibinfo{author}{\bibfnamefont{S.}~\bibnamefont{Bl\"ugel}},
  \bibinfo{journal}{Phys. Rev. B} \textbf{\bibinfo{volume}{90}},
  \bibinfo{pages}{115427} (\bibinfo{year}{2014}).

\bibitem[{\citenamefont{Draaisma and de~Jonge}(1988)}]{Draaisma:88.1}
\bibinfo{author}{\bibfnamefont{H.~J.~G.} \bibnamefont{Draaisma}}
  \bibnamefont{and} \bibinfo{author}{\bibfnamefont{W.~J.~M.}
  \bibnamefont{de~Jonge}}, \bibinfo{journal}{J. Appl. Phys.}
  \textbf{\bibinfo{volume}{64}}, \bibinfo{pages}{3610} (\bibinfo{year}{1988}).

\bibitem[{\citenamefont{Kubetzka et~al.}(2017)\citenamefont{Kubetzka, Hanneken,
  Wiesendanger, and von Bergmann}}]{Kubetzka:17.1}
\bibinfo{author}{\bibfnamefont{A.}~\bibnamefont{Kubetzka}},
  \bibinfo{author}{\bibfnamefont{C.}~\bibnamefont{Hanneken}},
  \bibinfo{author}{\bibfnamefont{R.}~\bibnamefont{Wiesendanger}},
  \bibnamefont{and} \bibinfo{author}{\bibfnamefont{K.}~\bibnamefont{von
  Bergmann}}, \bibinfo{journal}{Phys. Rev. B} \textbf{\bibinfo{volume}{95}},
  \bibinfo{pages}{104433} (\bibinfo{year}{2017}).

\bibitem[{\citenamefont{Dup\'e et~al.}(2016{\natexlab{a}})\citenamefont{Dup\'e,
  Bihlmayer, B\"ottcher, Bl\"ugel, and Heinze}}]{Dupe:16.1}
\bibinfo{author}{\bibfnamefont{B.}~\bibnamefont{Dup\'e}},
  \bibinfo{author}{\bibfnamefont{G.}~\bibnamefont{Bihlmayer}},
  \bibinfo{author}{\bibfnamefont{M.}~\bibnamefont{B\"ottcher}},
  \bibinfo{author}{\bibfnamefont{S.}~\bibnamefont{Bl\"ugel}}, \bibnamefont{and}
  \bibinfo{author}{\bibfnamefont{S.}~\bibnamefont{Heinze}},
  \bibinfo{journal}{Nature Comm.} \textbf{\bibinfo{volume}{7}},
  \bibinfo{pages}{11779} (\bibinfo{year}{2016}{\natexlab{a}}).

\bibitem[{\citenamefont{Bogdanov and
  Hubert}(1994{\natexlab{b}})}]{Bogdanov-1994aa}
\bibinfo{author}{\bibfnamefont{A.~N.} \bibnamefont{Bogdanov}} \bibnamefont{and}
  \bibinfo{author}{\bibfnamefont{A.}~\bibnamefont{Hubert}},
  \bibinfo{journal}{physica status solidi} \textbf{\bibinfo{volume}{186}},
  \bibinfo{pages}{527} (\bibinfo{year}{1994}{\natexlab{b}}).

\bibitem[{\citenamefont{Okubo et~al.}(2012)\citenamefont{Okubo, Chung, and
  Kawamure}}]{Okubo:12.1}
\bibinfo{author}{\bibfnamefont{T.}~\bibnamefont{Okubo}},
  \bibinfo{author}{\bibfnamefont{S.}~\bibnamefont{Chung}}, \bibnamefont{and}
  \bibinfo{author}{\bibfnamefont{H.}~\bibnamefont{Kawamure}},
  \bibinfo{journal}{Phys. Rev. Lett.} \textbf{\bibinfo{volume}{108}},
  \bibinfo{pages}{017206} (\bibinfo{year}{2012}).

\bibitem[{\citenamefont{Leonov and Mostovoy}(2015)}]{Leonov:15.1}
\bibinfo{author}{\bibfnamefont{A.~O.} \bibnamefont{Leonov}} \bibnamefont{and}
  \bibinfo{author}{\bibfnamefont{M.}~\bibnamefont{Mostovoy}},
  \bibinfo{journal}{Nature Comm.} \textbf{\bibinfo{volume}{6}},
  \bibinfo{pages}{8275} (\bibinfo{year}{2015}).

\bibitem[{\citenamefont{Dup\'e et~al.}(2016{\natexlab{b}})\citenamefont{Dup\'e,
  Kruse, Dornheim, and Heinze}}]{Dupe:16.2}
\bibinfo{author}{\bibfnamefont{B.}~\bibnamefont{Dup\'e}},
  \bibinfo{author}{\bibfnamefont{C.~N.} \bibnamefont{Kruse}},
  \bibinfo{author}{\bibfnamefont{T.}~\bibnamefont{Dornheim}}, \bibnamefont{and}
  \bibinfo{author}{\bibfnamefont{S.}~\bibnamefont{Heinze}},
  \bibinfo{journal}{New J. Phys.} \textbf{\bibinfo{volume}{18}},
  \bibinfo{pages}{055015} (\bibinfo{year}{2016}{\natexlab{b}}).

\bibitem[{\citenamefont{R\'ozsa et~al.}(2017)\citenamefont{R\'ozsa, Palot\'as,
  De\'ak, Simon, Yanes, Udvardi, Szunyogh, and Nowak}}]{Rosza:17.1}
\bibinfo{author}{\bibfnamefont{L.}~\bibnamefont{R\'ozsa}},
  \bibinfo{author}{\bibfnamefont{K.}~\bibnamefont{Palot\'as}},
  \bibinfo{author}{\bibfnamefont{A.}~\bibnamefont{De\'ak}},
  \bibinfo{author}{\bibfnamefont{E.}~\bibnamefont{Simon}},
  \bibinfo{author}{\bibfnamefont{R.}~\bibnamefont{Yanes}},
  \bibinfo{author}{\bibfnamefont{L.}~\bibnamefont{Udvardi}},
  \bibinfo{author}{\bibfnamefont{L.}~\bibnamefont{Szunyogh}}, \bibnamefont{and}
  \bibinfo{author}{\bibfnamefont{U.}~\bibnamefont{Nowak}},
  \bibinfo{journal}{Phys. Rev. B} \textbf{\bibinfo{volume}{95}},
  \bibinfo{pages}{094423} (\bibinfo{year}{2017}).

\bibitem[{\citenamefont{Bessarab et~al.}(2012)\citenamefont{Bessarab, Uzdin,
  and J\'onsson}}]{Bessarab:12.1}
\bibinfo{author}{\bibfnamefont{P.~F.} \bibnamefont{Bessarab}},
  \bibinfo{author}{\bibfnamefont{V.~M.} \bibnamefont{Uzdin}}, \bibnamefont{and}
  \bibinfo{author}{\bibfnamefont{H.}~\bibnamefont{J\'onsson}},
  \bibinfo{journal}{Phys. Rev. B} \textbf{\bibinfo{volume}{85}},
  \bibinfo{pages}{184409} (\bibinfo{year}{2012}).

\bibitem[{\citenamefont{R\'ozsa et~al.}(2016)\citenamefont{R\'ozsa, Simon,
  Palot\'as, Udvardi, and Szunyogh}}]{Rosza:16.1}
\bibinfo{author}{\bibfnamefont{L.}~\bibnamefont{R\'ozsa}},
  \bibinfo{author}{\bibfnamefont{E.}~\bibnamefont{Simon}},
  \bibinfo{author}{\bibfnamefont{K.}~\bibnamefont{Palot\'as}},
  \bibinfo{author}{\bibfnamefont{L.}~\bibnamefont{Udvardi}}, \bibnamefont{and}
  \bibinfo{author}{\bibfnamefont{L.}~\bibnamefont{Szunyogh}},
  \bibinfo{journal}{Phys. Rev. B} \textbf{\bibinfo{volume}{93}},
  \bibinfo{pages}{024417} (\bibinfo{year}{2016}).

\bibitem[{\citenamefont{Vosko et~al.}(1980)\citenamefont{Vosko, Wilk, and
  Nusair}}]{doi-10.1139/p80-159}
\bibinfo{author}{\bibfnamefont{S.~H.} \bibnamefont{Vosko}},
  \bibinfo{author}{\bibfnamefont{L.}~\bibnamefont{Wilk}}, \bibnamefont{and}
  \bibinfo{author}{\bibfnamefont{M.}~\bibnamefont{Nusair}},
  \bibinfo{journal}{Canadian Journal of Physics} \textbf{\bibinfo{volume}{58}},
  \bibinfo{pages}{1200} (\bibinfo{year}{1980}).

\bibitem[{\citenamefont{Mentink et~al.}(2010)\citenamefont{Mentink, Tretyakov,
  Fasolino, Katsnelson, and Rasing}}]{Mentink2010}
\bibinfo{author}{\bibfnamefont{J.~H.} \bibnamefont{Mentink}},
  \bibinfo{author}{\bibfnamefont{M.~V.} \bibnamefont{Tretyakov}},
  \bibinfo{author}{\bibfnamefont{A.}~\bibnamefont{Fasolino}},
  \bibinfo{author}{\bibfnamefont{M.~I.} \bibnamefont{Katsnelson}},
  \bibnamefont{and} \bibinfo{author}{\bibfnamefont{T.}~\bibnamefont{Rasing}},
  \bibinfo{journal}{J. Phys. C} \textbf{\bibinfo{volume}{22}},
  \bibinfo{pages}{176001} (\bibinfo{year}{2010}).

\end{thebibliography}

\end{document}